\documentclass[a4paper,fleqn]{cas-dc}

\usepackage[numbers]{natbib}
\usepackage{subfigure}
\usepackage{lineno,hyperref}
\usepackage{graphicx,floatrow}
\usepackage{subfigure}
\usepackage{amsfonts}
\usepackage{amssymb}
\usepackage{amsmath}
\usepackage{algorithm}
\usepackage{algorithmic}
\usepackage{makecell,rotating,multirow,diagbox}
\usepackage{epstopdf}
\usepackage{color}
\usepackage{textcomp}
\usepackage{bm}
\usepackage{threeparttable}
\usepackage{booktabs}
\usepackage{graphicx}
\usepackage{longtable}
\usepackage{booktabs}
\usepackage{supertabular}
\usepackage{pdfpages}
\usepackage{calc}
\usepackage{url}
\usepackage{tikz}
\usepackage{lettrine}
\usepackage{epstopdf}
\usepackage{graphicx} 

\begin{document}
\let\WriteBookmarks\relax
\def\floatpagepagefraction{1}
\def\textpagefraction{.001}
\shorttitle{A Trustless Architecture of Blockchain-enabled Metaverse}
\shortauthors{Minghui Xu et~al.}

\title [mode = title]{A Trustless Architecture of Blockchain-enabled Metaverse}
\author[1]{Minghui Xu}[orcid=0000-0003-3675-3461]
\ead{mhxu@sdu.edu.cn}
\author[1]{Yihao Guo}[orcid=0000-0003-3266-6002]
\ead{yhguo@mail.sdu.edu.cn}
\author[2]{Qin Hu}[orcid=0000-0002-8847-8345]
\ead{qinhu@iu.edu}
\author[3]{Zehui Xiong}[orcid=0000-0002-4440-941X]
\ead{zehui_xiong@sutd.edu.sg}
\author[1]{Dongxiao Yu}[orcid=0000-0001-6835-5981]
\ead{dxyu@sdu.edu.cn}
\author[1]{Xiuzhen Cheng}[orcid=0000-0001-5912-4647]
\ead{xzcheng@sdu.edu.cn}
\cormark[1]
\cortext[1]{Corresponding author}
\address[1]{School Computer Science and Technology, Shandong University, Qingdao, China}
\address[2]{Department of Computer and Information Science, Indiana University-Purdue University Indianapolis, USA}
\address[3]{Information Systems Technology and Design Pillar, Singapore University of Technology and Design, Singapore}

\begin{abstract}
Metaverse has rekindled human beings' desire to further break space-time barriers by fusing the virtual and real worlds. However, security and privacy threats hinder us from building a utopia. A metaverse embraces various techniques, while at the same time inheriting their pitfalls and thus exposing large attack surfaces. Blockchain, proposed in 2008, was regarded as a key building block of metaverses. it enables transparent and trusted computing environments using tamper-resistant decentralized ledgers. Currently, blockchain supports Decentralized Finance (DeFi) and Non-fungible Tokens (NFT) for metaverses. However, the power of a blockchain has not been sufficiently exploited. In this article, we propose a novel trustless architecture of blockchain-enabled metaverse, aiming to provide efficient resource integration and allocation by consolidating hardware and software components. To realize our design objectives, we provide an On-Demand Trusted Computing Environment (OTCE) technique based on local trust evaluation. Specifically, the architecture adopts a hypergraph to represent a metaverse, in which each hyperedge links a group of users with certain relationship. Then the trust level of each user group can be evaluated based on graph analytics techniques. Based on the trust value, each group can determine its security plan on demand, free from interference by irrelevant nodes. Besides, OTCEs enable large-scale and flexible application environments (sandboxes) while preserving a strong security guarantee. 
\end{abstract}


\maketitle

\section{Introduction}
The concept of metaverse was originated from Snow Cash, a 1992 science fiction novel by Neal Stephenson. The word ``metaverse'' is a portmanteau of ``meta'' (meaning transcending) and ``verse'' (abbreviation of the universe). In 2021, metaverse became popular overnight since it rekindled people's hope of building an ideal virtual society where human beings are tightly connected. Big companies then started to commit to developing metaverse software, 
e.g., Meta Horizon Workroom~\cite{HorizonWorkroom}, Microsoft Mesh~\cite{MicrosoftMesh}, and NVIDIA Omniverse~\cite{Omniverse}. In fact, human beings have undergone a long history of building tight bonds and shortening the distance among themselves, from ancient times to the current information age. Nevertheless, constructing a metaverse is a challenging task, though exciting and stirring. 

Before metaverse is pushed to forefront, many efforts have been made, including Virtual Reality (VR)/Augmented Reality (AR), 3D virtual world, and online video games~\cite{dionisio20133d}. 
However, how to implement a true metaverse is still unclear and controversial to developers, even though they have constructed a number of relevant tools and established a common goal of fusing virtual and reality. The major concerns about metaverse are related to its feasibility and safety since a metaverse can consume a tremendous amount of computational resources and require a safe and trusted environment. 
Based on our investigations, we summarize the key issues of metaverse as follows. 

\begin{figure}[!htbp]
\centering
\includegraphics[width=0.9\textwidth]{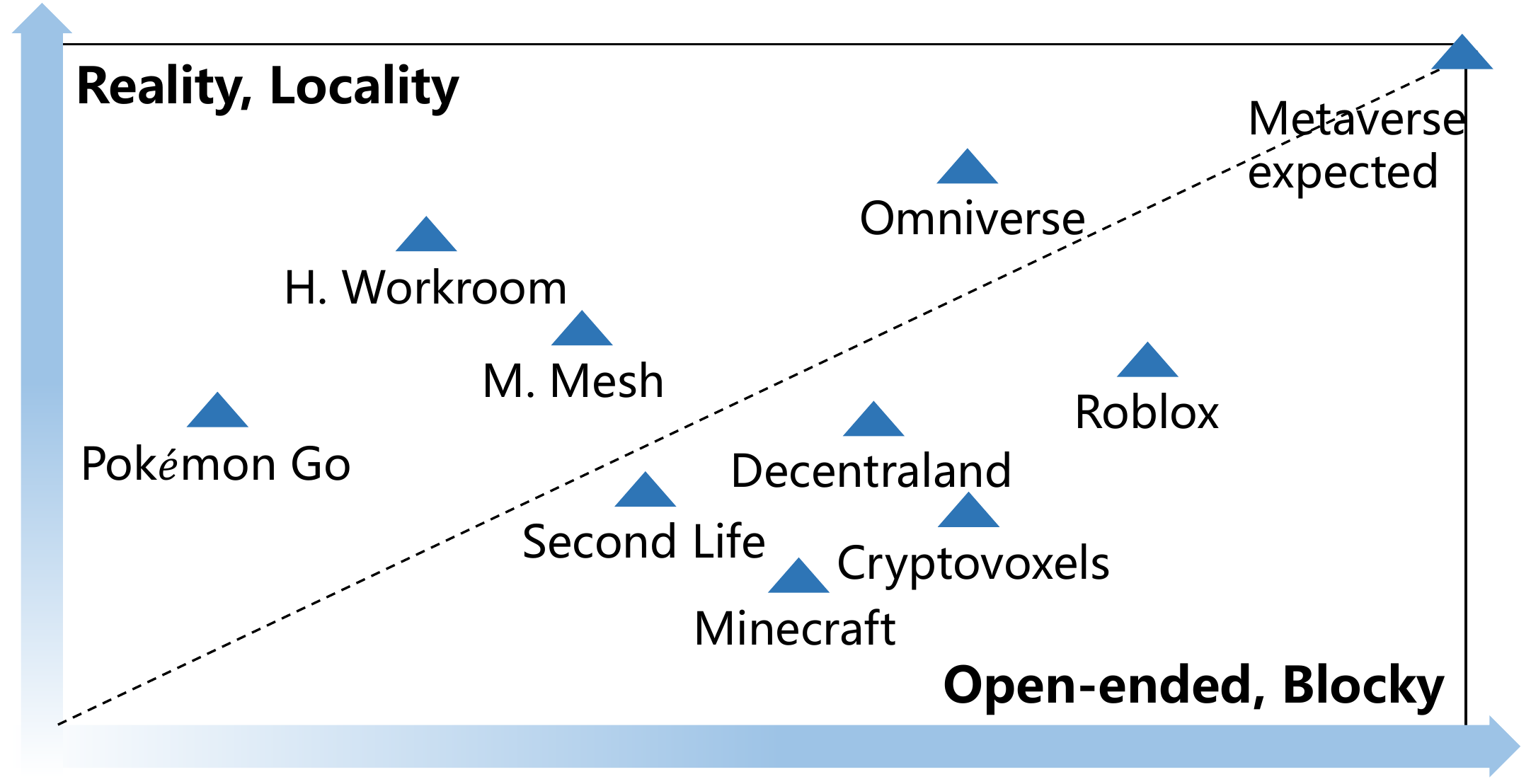}
\caption{The roadmap of the metaverse projects.
}
\label{fig:roadmap}
\end{figure}

First, current infrastructures can hardly support building a metaverse that meets our expectations. Metaverse applications currently exhibit a tradeoff between reality and openness as shown in Figure~\ref{fig:roadmap}. Resource shortage and improper allocation lead to the sacrifice of either openness or reality. For example, the resource in Minecraft~\cite{Minecraft} is mainly used to improve the openness for users but sacrifices the reality (users can only live in a ``blocky'' world). Hence there is an urgent demand for an efficient metaverse architecture, which can sufficiently utilize the existing computational resources.
Second, security and privacy threats hinder the building of a practical metaverse platform. For example, at the beginning of 2022, the metaverse company Meta was sued for illegally collecting facial information without users' consent~\cite{Meta}. Besides, the attack surface of a metaverse is very largely due to its diversity \cite{wang2022survey}. Additionally, a metaverse needs to integrate various technologies whose pitfalls are also naturally inherited. For instance, blockchain, proposed in 2008, has been regarded as the key building block of a metaverse. Blockchain enables transparent and trusted computing environments using a tamper-resistant decentralized ledger; however, the functions of a blockchain are limited, as one can see that it is only used to build Decentralized Finance (DeFi) or Non-Fungible Token (NFT) nowadays. Besides, using blockchain in a metaverse can consume a large amount of resources and incur long latency. 

In this paper, we attempt to address the aforementioned shortcomings with a novel metaverse architecture. Our contributions are highlighted as follows:

\begin{itemize}
\item We propose an architecture of blockchain-based metaverse by consolidating hardware and software components, aiming at providing efficient resource integration and allocation. The architecture presents detailed collaborations among different modules.  
\item We formally define a local trust model (LTM) to depict a metaverse as a weighted hypergraph. Using this model, it is feasible to evaluate the trust among each group of metaverse users and provide it with an appropriate computing environment according to the trust value. 
\item To enhance security and privacy of a metaverse, we propose On-Demand Trusted Computing Environment (OTCE) to support application environments with variable scalability, which can provide a strong security guarantee using blockchain as an underlying technology.
\end{itemize}

\section{Related Works and Motivation}

Since metaverse is a young field, we summarize the effort from both industry and academia made in the past two years. 

\subsection{Metaverse in Industry}

Facebook officially announced that its company name was changed to Meta~\cite{Meta} in 2021, which marks that it has identified metaverse as an important direction for future. 
Horizon Workroom~\cite{HorizonWorkroom}, launched by Meta, can provide people with immersive virtual meeting rooms. Users in any physical location with an Oculus Quest 2 helmet can join a Horizon Workroom and experience virtual whiteboard control and file sharing.
Microsoft Mesh~\cite{MicrosoftMesh} is a platform supported by Azure. It adopts technologies including blockchain, artificial intelligence (AI), and extended reality (XR) to accomplish virtual collaborations and spatially aware design reviews. 
NVIDIA presented Omniverse~\cite{Omniverse}, which is a metaverse system focusing on 3D simulation and digital twins. 
The Omniverse ecosystem includes components such as Omniverse Connect, Nucleus, and RTX, among which RTX can provide Omniverse with powerful computing powers.

In the field of 3D games, Roblox~\cite{Roblox} proposed eight key characteristics of a metaverse, i.e., identity, friends, immersion, anywhere, diversity, low latency, economy, and civilization. 
Second Life~\cite{secondlife} is a free 3D virtual world where users can create and connect with others using voice and text. Linden Dollar as the virtual currency of Second Life can be exchangeable with real-world currency.
Minecraft~\cite{Minecraft} is an online 3D game that allows users to create virtual worlds based on their creativity. 
Users can enter a game through VR devices, e.g., Oculus Rift, thereby enhancing the sense of immersion. 
Pok\'{e}mon Go~\cite{Pokemon} is a location-based AR game, in which players can use mobile devices (iPhone and Android devices) to travel between the real world and the virtual world. 

Some projects introduce blockchain to the metaverses they build. Decentraland is a decentralized VR platform based on Ethereum, in which users can obtain benefits by creating, experiencing, and developing NFT. 
Cryptovoxels~\cite{Cryptovoxels} is a virtual world built on Ethereum where players can buy, sell and construct virtual art galleries, shops, etc. 

\subsection{Metaverse in Academia}

Jon Radoff~\cite{Jon2021metaverse} presented a metaverse architecture with the following seven layers: experience, discovery, creator economy, spatial computing, decentralization, human interface, and infrastructure.
CUHKSZ~\cite{duan2021metaverse} is a university campus prototype implemented with the FISCO-BCOS consortium blockchain. CUHKSZ supports tokens, Distributed Autonomous Organizations (DAO), and trading. The creators also put forward a number of critical challenges and thoughtful questions about developing a metaverse. 
Van~\textit{et al.}~\cite{van2022edge} proposed a new digital twin scheme, which adopts mobile edge computing (MEC) and ultra-reliable and low latency communications (URLLC) technologies to help metaverse improve reliability and reduce communication latency.
Nair~\textit{et al.}~\cite{nair2022going} presented an $\varepsilon$-differential privacy framework to improve the security and privacy of VR devices, which enables users to maximize privacy while minimizing usability impact when using VR devices to participate in the metaverse. 
In \cite{nair2022exploring}, Nair~\textit{et al.} further explored privacy risks in the metaverse through an experiment with 30 researchers.

In addition to the works mentioned above, there also exist a few surveys and reviews, which intend to summarize the effort made toward metaverse from various perspectives.
Dionisio~\textit{et al.}~\cite{dionisio20133d} pointed out four directions of developing a metaverse, namely immersive realism, ubiquity of access and identity, interoperability, and scalability. 
Kye~\textit{et al.}~\cite{kye2021educational} discussed the opportunities and challenges of metaverse in education.
Damar~\textit{et al.}~\cite{damar2021metaverse} extracted relevant information about the development of metaverse from the past three decades.
Park~\textit{et al.}\cite{park2022metaverse} classified and analyzed the current metaverse schemes from five perspectives, i.e., hardware, software, contents, user interaction, implementations, and applications.
In~\cite{wang2022survey}, Wang~\textit{et al.} 
investigated the problems of the current major metaverse solutions from the perspective of security and privacy.
The focus of Xu~\textit{et al.}\cite{xu2022full} is on an edge-enabled metaverse. 
Yang~\textit{et al.}\cite{yang2022fusing}
studied the important role that AI and blockchain play in metaverse.

\subsection{Motivation}
According to the above description on the most related works, one can see that metaverse in industry is still in its exploration stage. Companies are exploring the possibilities of building metaverses in various application scenarios such as 3D games, NFT, online collaborations, and digital twins.
However, as shown in Figure~\ref{fig:roadmap}, current metaverse projects cannot sufficiently balance reality and openness. For example, some projects, including Second Life, Minecraft, and Roblox, have good openness and allow users to create their worlds according to their own interests; but it is difficult for these projects to guarantee reality, and the users can only employ limited tools, such as pixel squares in Minecraft, to piece together things in a ``blocky'' world. Other projects, including Pok\'{e}mon Go, Horizon Workroom, and Omniverse, can achieve reality but their virtual worlds are not fully opened. In academia, most papers are reviews and surveys, which mainly summarize the major metaverse platforms and propose prospects for future metaverse developments. A few articles propose layered architectures but lack sufficient and specific design details. Besides, rational allocation of resources, security, and privacy issues faced by metaverses are largely overlooked. 

However, a metaverse is by no means a simple combination of technologies, but in need of considering a reasonable consolidation of hardware and software, so as to realize the efficient integration and allocation of resources. Only with sufficient resources can one combine reality and openness. Moreover, metaverse should take security and privacy into its initial design. As one of the key technologies to build a metaverse, blockchain~\cite{belotti2019vademecum, kolb2020core} can create a trusted interactive environment for users who do not trust each other and has the potential to be used as the underlying technology to provide security guarantee for the adoption of other technologies. However, current metaverse designs do not fully explore the potential of blockchain, which should have played a more important role in a metaverse~\cite{yang2022fusing}. Motivated by these considerations, we propose a novel trustless metaverse architecture enabled by blockchain in this paper, which realizes a reasonable allocation of resources and improves the security and privacy of a metaverse.

\section{A Trustless Architecture of Blockchain-enabled Metaverse}

\subsection{Architecture}

\subsubsection{Resource Integration}
A metaverse is definitely a large-scale data-intensive system. Nevertheless, current computing technologies are not sufficiently advanced to satisfy metaverses' huge computational power requirements, making developers have to sacrifice either reality or openness. High-performance hardware (especially Graphics Processing Units (GPUs)~\cite{brodtkorb2013graphics}) can tackle complex rendering workloads and thus make a small virtual scenario approach to reality. However, it is almost impossible to render every part of a metaverse due to resource shortage. Therefore, we have to sacrifice reality to make the metaverse open-ended, or vice versa. Overall, the current networking and computing technologies are the bottlenecks of realizing a metaverse as we expect. The fundamental challenge is: how to address the exponentially increasing number of sensors deployed for a metaverse and amount of data collected by the sensors with limited hardware support? 

Therefore, we propose a new architecture by fusing blockchain and edge computing technologies to integrate all the computational resources from cloud-edge-device layers. A blockchain serves as an underlying trusted intermediary, which connects all the resources within a large network. Similar to Sky computing~\cite{keahey2009sky}, our blockchain intends to relate all the isolated services provided by large and small companies, so as to aggregate the resources we can have for building a metaverse. Besides, the network is virtualized as a resource pool, in which computation tasks are assigned to computing units based on utilization conditions. By this way, the integration of blockchain and various computational resources yields a virtual giant computer, and one can maximize resource utilization with all hardware in hands. 

\begin{figure*}[!htbp]
\centering
\centerline{\includegraphics[width=\textwidth]{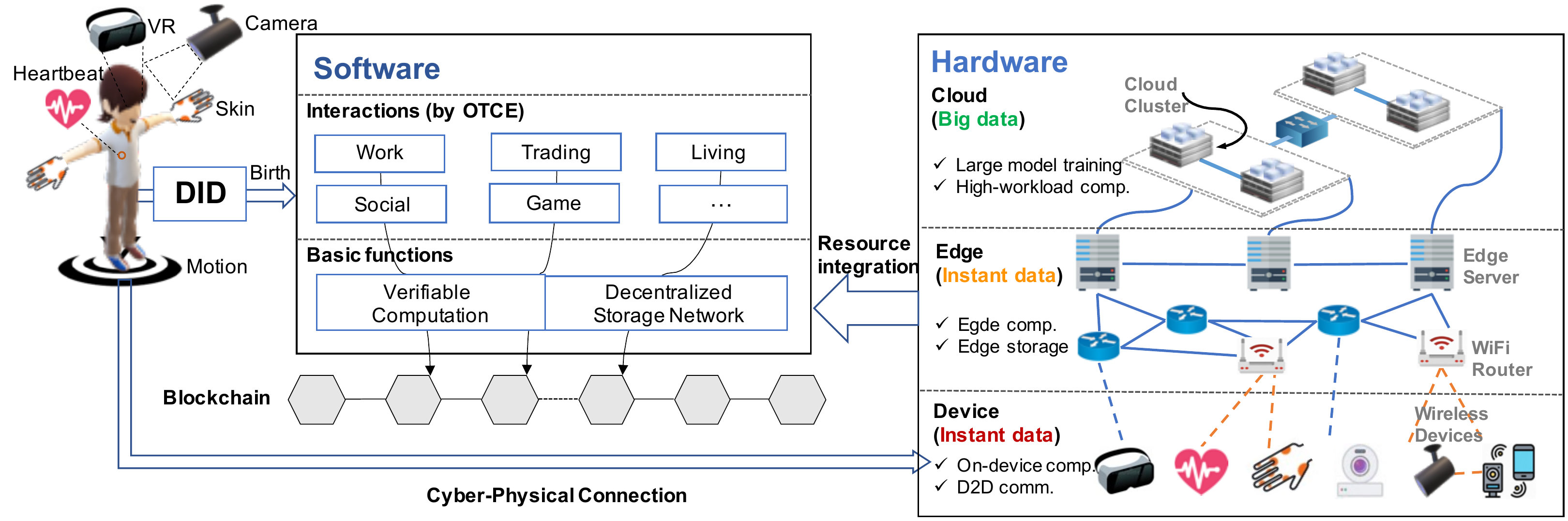}}
\caption{ A metaverse is established on the basis of software and hardware infrastructures. Each participant is surrounded by a variety of sensors which connect the physical and cyber worlds. OTCE controls interactions among digital identities, which are supported by two basic functions. All software components are safeguarded by a blockchain. The hardware is coordinated to contribute resources and is optimized under the goal of realtimeness.}
\label{fig:arch}
\end{figure*}
\subsubsection{Overview}
As shown in Figure \ref{fig:arch}, the architecture of a metaverse consists of two major components: software and hardware.
A real person is physically surrounded by a variety of sensors such as a heartbeat sensor, a VR headset, a camera, a skin sensor, and a motion sensor. These sensors connect the physical and cyber worlds. A virtual human must obtain a decentralized identity (DID) before being given birth in the metaverse.

In hardware, resource allocation follows the following rule: maximize user experience with ultra-low latency. 
Cloud servers are commonly used to train large AI models and carry out high-workload computations. This layer works for big data processing. The edge layer and device layer deal with instant data. Edge servers form a backbone to carry out basic functions including verifiable computation and decentralized storage, contributing to computation and storage services. Personal devices at the edge layer can participate as service providers. People can offer resources to share and get rewards. By this way we are able to gather enough resources for the metaverse. The device layer deals with instant data that is hotter than those at the edge layer. On-device computation is an efficient way when tackling real-time applications, such as processing videos, tracking objects, and sending alarms. With specific hardware support (e.g., GPU, Tensor Processing Unit (TPU)~\cite{jouppi2017datacenter}), we can perform on-device learning directly on a local device. Moreover, IoT devices are resource-constrained in computation, storage, and energy, but they can still reach consensus and collaborate through message passing (communication-heavy). 
In summary, based on the above design, the Hardware can allocate the corresponding computing resources according to the user's demand for the timeliness of data processing and maximize user experience with ultra-low latency. Due to the incentive mechanism in the device layer, the device owners around the requester would actively provide their  computing power to help requesters complete the computational tasks and get rewards. Therefore, some hotter instant data would be processed immediately by nearby devices, providing low-latency services while ensuring a good user experience.

In software, we introduce two novel approaches, LTM and OTCE, which make the metaverse system controllable and secure. There are two basic functions underlying the computing and storage infrastructure, both being supported by a blockchain: verifiable computation ensures that the computing process is fault-tolerant and the computational results are verifiable; while a decentralized storage network distributes voluminous data into different types of storage nodes. LTM measures the trust value of a group of nodes. Based on the trust value, OTCEs provide trusted services for each group that can maintain diversified applications in the metaverse. OTCEs are built on top of the two basic functions and can be regarded as ``sandboxes''. 

\subsection{Distributed Identity}
Distributed Identity (DID) is a critical technique applied in metaverse. A DID represents a legitimate identifier and verifiable credential for a metaverse participant. Usually, each identity is mapped to an avatar. For example, CryptoAvatars built on top of an Ethereum blockchain creates 3D avatars in the form of NFTs. Tokenized avatars can be exchanged or transferred in a trustless platform. Note that avatars are not necessary in some cases where users can directly interact with each other without relying on a virtual character. For example, when Alice and Bob are trading digital commodities, they can issue transactions without creating avatars first. 

However, it remains a challenging problem that how to guarantee a secure and precise mapping from an identity or avatar to a real person. Such a capability is indispensable when establishing a metaverse. Without a secure and reliable mapping technique, a malicious user can launch Sybil attacks~\cite{douceur2002sybil}. For example, an attacker can control an election process using numerous Zombie accounts. 

Hence we propose a new method of constructing DIDs with a strong security guarantee. Our method focuses on the mapping from digital identity to physical identity and supports large-scale identity establishments in harsh environments. On one hand, users' personal information is collected by sensors and transferred to a DID attestation module. The DID attestation module is built with smart contracts, by which all the formats and policies are verified in a trustless and automatic way. To achieve precise attestation, we can deploy biometric technologies such as fingerprint and facial recognition~\cite{unar2014review}. On the other hand, DID supports asynchronous distributed key generation (ADKG)~\cite{kokoris2020asynchronous} apart from the public key infrastructure. Asynchronous byzantine algorithms should work under the assumptions of both asynchronous networks and Byzantine nodes. An asynchronous network refers to the case when the message delay can be infinitely long. Byzantine nodes can behave arbitrarily such as crash or collude. ADKG distributes a secret among a group. A sufficient number of nodes can later reconstruct the secret by combining their shares when they want to use the secret in tasks such as group authentication, byzantine agreement, and multi-party computation. 










\subsection{Trust Evaluation}

Trust is the basis of collaboration. A trust model can output a trust value influenced by various factors such as network size, message latency, and node behaviors. In a metaverse, applying a global trust model is infeasible since one model can hardly depict the complex relationships among all users, both virtual and real. Therefore, our architecture adopts a local trust model (LTM). LTM satisfies a \textit{locality} property, which means that we can measure the local trust with only a subset of nodes. In comparison, a global trust model implies that trust is evenly distributed. For example, the upper bound $f=N/3$ about Byzantine nodes usually refers to the case in which any node might become Byzantine with equal probability but no more than $N/3$ nodes are Byzantine at the same time in total. Byzantine-resilient protocols can have strong security guarantees when confronted with $f$ malicious nodes. In such a circumstance, a subset of nodes with high trust values has to follow the protocol without exception and thus endure the unnecessary high communication overheads. Therefore, a global trust model fits into simple distributed systems such as a blockchain or a P2P network. However, global models do not suit a complicated metaverse because treating all nodes alike ignores the differences between different subsets of nodes, thus negatively impacting the system performance and user experience.

Essentially, LTM assigns each group of nodes a trust value to depict trust in a fine-grained manner, rather than offer a unified trust value for the entire network. LTM provides metaverse networks with a rigorous mathematical expression, from which one can benefit greatly when setting up a trusted computing environment or carrying out tasks such as social mining. 

\begin{figure}[!htbp]
\centering
\centerline{\includegraphics[width=0.5\textwidth]{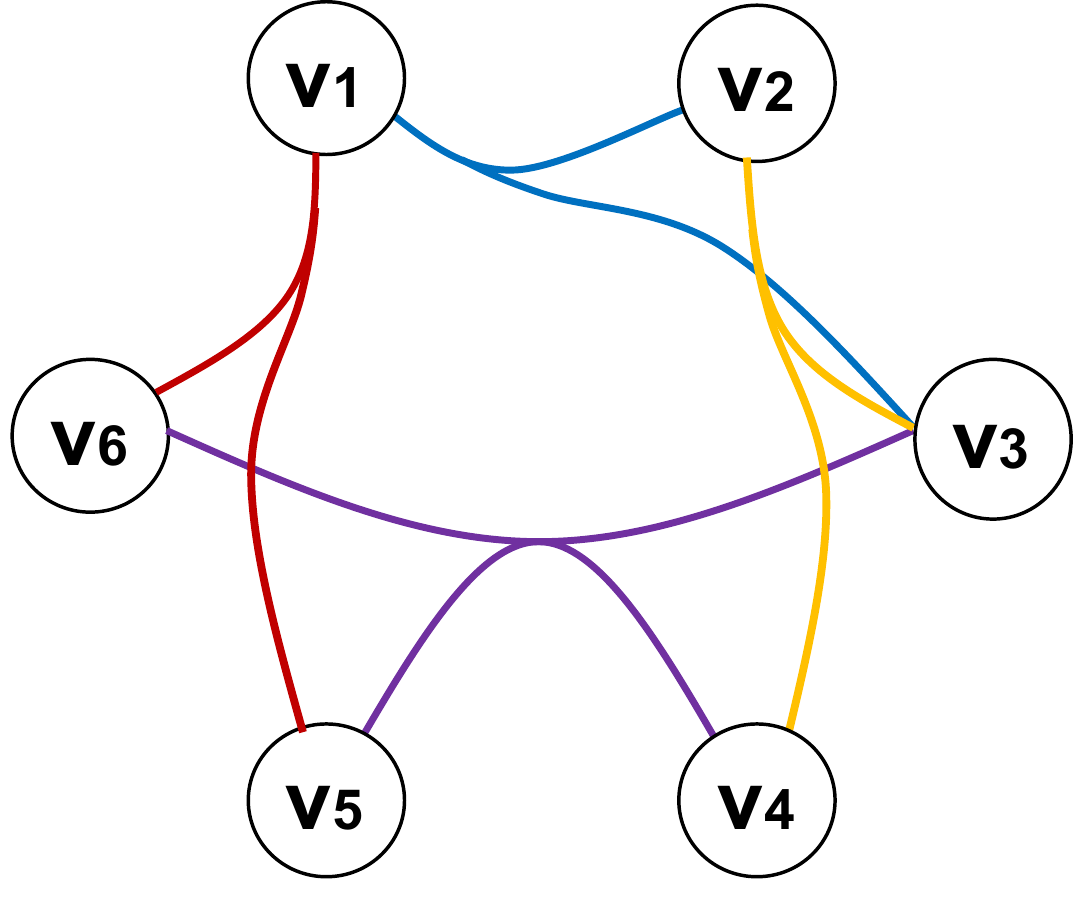}}
\caption{LTM as a hypergraph}
\label{fig:hypergraph}
\end{figure}

To have a better illustration on LTM, we provide a symbolized expression of an LTM as a hypergraph~\cite{bretto2013hypergraph} shown in Figure~\ref{fig:hypergraph}. Specifically, we denote a metaverse network as a weighted hypergraph $H=(V, E, W)$, where $V$ is the set of nodes, $E$ is the set of hyperedges, and $W$ is the set of weights. Each hyperedge $e\in E$ represents a relation among a group of $k$ nodes $\{v_1, v_2, \cdots, v_k \}$, denoted by  $e=\{v_1, v_2, \cdots, v_k \}$, and is weighted by a \textit{trust value} depicting to what extent the nodes in the group trust each other. In the metaverse, a pairwise relationship is insufficient to depict social connections. Hypergraph fits the multi-adic relationships well, and each group of nodes collaborates to establish their own LTM. Thus, one can quantify the metaverse network and study its characteristics by various graph analytics techniques. For example, we can estimate the distribution of computing power usage from the degree distribution and clustering coefficients. The estimation can be further used to determine the resource allocation of hardware. 
Using LTMs and hypergraphs, we can create sandboxes for massive interacted applications with strong security guarantees and formulate the metaverse in a formal and computable graph model. The metaverse hereby becomes controllable and auditable. 

Generally, a metaverse can aggregate the sensed network data to calculate trust. However, poor data quality can lead to inaccurate trust estimation. Data quality is measured based on factors of completeness, consistency, and reliability. Hence we introduce Oracle to serve as a data transport module and provide high-quality data. An oracle (e.g., Provable) is a third-party trusted service that collects data from the digital world. It can provide authenticity proofs to demonstrate that the data source is authentic and untampered. Relying on oracles, we can evaluate trust more accurately and efficiently. 
Note that, currently, oracles are commonly connected to blockchain systems, which improves the security of the data flowing into the blockchain and is considered one of the best solutions for the smart contract to compute based on inputs from the real world~\cite{Oracle, pasdar2022connect}. Moreover, as a third party, the oracle does not affect the blockchain consensus and other processes, so the entire system is still decentralized.


\subsection{On-Demand Trusted Computing Environment}

The role a blockchain plays definitely is beyond cryptocurrency, DeFi, and NFT. Even though blockchains have demonstrated their effectiveness in enhancing the security of cloud~\cite{CloudChain}, edge~\cite{Curb, SPDL} and device ~\cite{BLOWN, wChain, TEMS} layers, their power has not been unleashed in a metaverse. In our metaverse architecture, blockchains are expected to help the metaverse build a trusted ecosystem. To meet this goal, we have to address three challenges. 
First, blockchains should be appropriately and lightly deployed in the metaverse to support massive data-intensive applications in parallel. Storing hashes of data chunks on a blockchain might incur unaffordable storage overheads in the metaverse even though this approach is effective in saving storage of traditional blockchain systems.
Secondly, even if the throughput of a blockchain has been dramatically improved (e.g., from 6 TPS to 100 kTPS) since 2008, deploying and terminating blockchains are still costly and time-consuming. In fact, it is hard for small companies or organizations to quickly deploy a blockchain in a short time due to the prohibitively high cost. 
Third, cross-chain interoperability becomes a bottleneck in the current blockchain ecosystem, in which more than ten thousand independent blockchain systems are isolated and heterogeneous; thus unifying blockchains deployed in a metaverse is very challenging.

\begin{figure}[!htbp]
\centering
\centerline{\includegraphics[width=\textwidth]{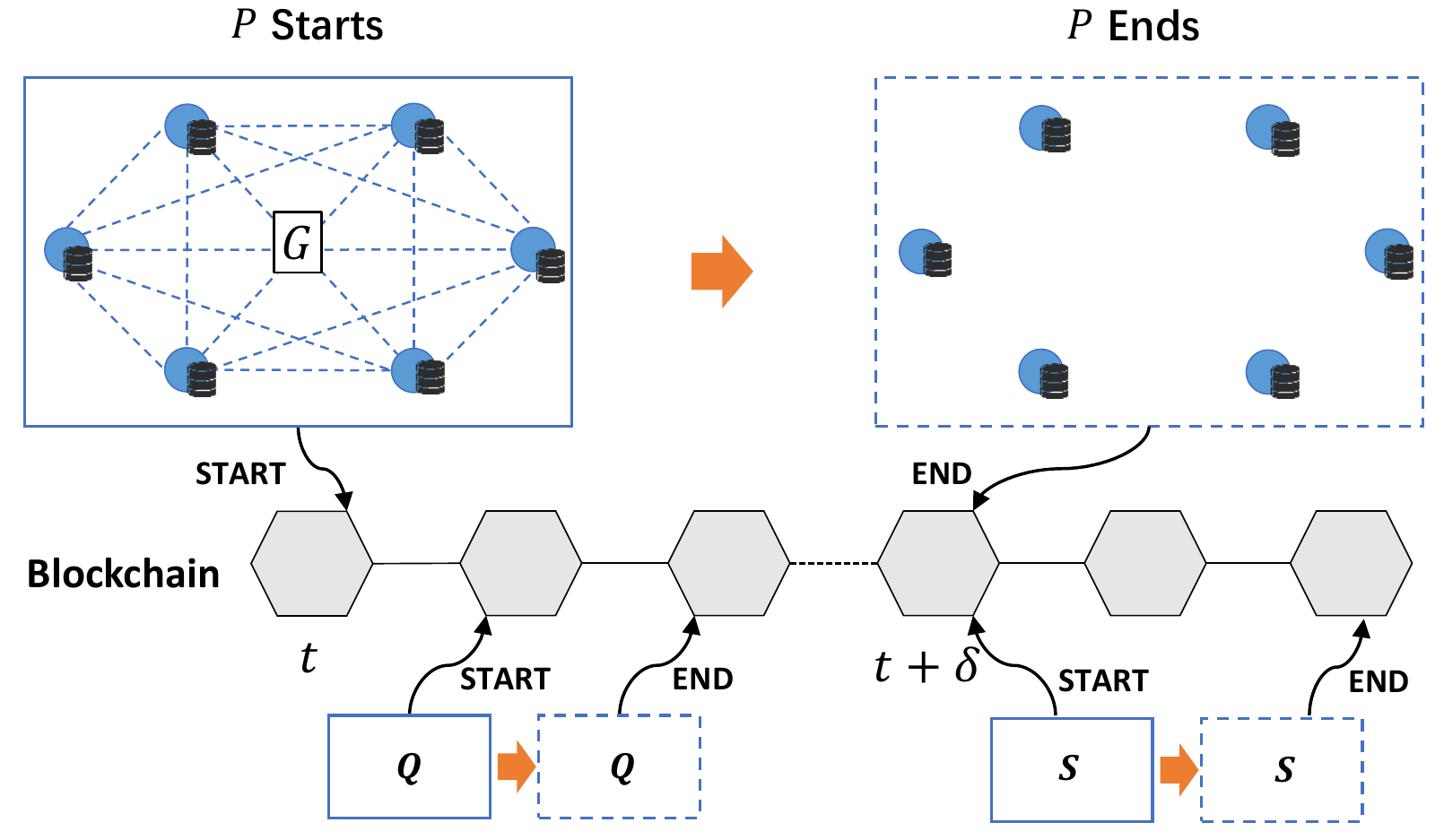}}
\caption{On-demand Trusted Computing Environment (OTCE)}
\label{fig:octe}
\end{figure}

To address the above challenges, we propose On-Demand Trusted Computing Environment (OTCE), which serves as an underlying infrastructure to provide trust throughout the metaverse. The design goals of an OTCE are listed as follows. 
\begin{itemize}
    \item \textbf{Security.} The lifecycle of each OTCE is protected. State transitions of an OTCE are fully recorded in a tamper-proof way. 
    \item \textbf{Flexibility.} An OTCE can be efficiently created, manipulated, and terminated in the metaverse. 
    \item \textbf{Lightweight.} Data flows are closely related to a specific OTCE. Different OTCEs can freely share real-time data to avoid storage redundancy. 
\end{itemize}

\subsubsection{OTCE Details} 
With OTCE, a group of nodes can collaboratively establish a trusted computing environment on demand. Recall that in on-chain computation, the entire computation process is executed by all miners via smart contracts, results are validated by validators (or miners), and confirmed results are recorded on a blockchain and then become readable to all nodes. In contrast, off-chain computation occurs outside of smart contracts with only critical information being written to the blockchain. 
An OTCE can be created when needed and then terminated when its duty is complete. Some predefined operations are required to be on-chain, but most of the computation processes are carried out off-chain. 
For example, Figure~\ref{fig:octe} presents three OTCEs ($P$, $Q$, and $S$) from opening to closing, and we take $P$ as an example to show the whole process of an OTCE. 
To be specific, multiple nodes form a group $G$ to request the blockchain to establish an on-demand trusted computing environment $P$ at time $t$, and the request is verified by the miners. 
When the verification is successful, the behavior of opening OTCE is recorded on the blockchain and the participating nodes can interact with each other in OTCE. 
When the node interaction in $P$ is completed (at time $t+\delta$), the participants in $G$ upload their final results based on the verifiable computing techniques, e.g., trusted hardware and zero-knowledge proofs~\cite{yu2017survey} (more descriptions are shown in Sec.~\ref{sec:BVM}), to the blockchain and apply to close $P$. After successful verification by miners, the blockchain closes the OTCE $P$.

Formally, we denote an OTCE by a vector $\Vec{\mathsf{E}}$=$(\mathsf{EID}$, $\mathsf{State}$, $\mathsf{G}$, $\Delta T$, $\mathsf{Results})$. Environment identity (abbreviated by $\mathsf{EID}$) is a unique identity of an OTCE, making OTCEs distinguishable and traceable. $\mathsf{State}$ represents the current state of $\Vec{\mathsf{E}}$. An OTCE has four different states, denoted by $\mathsf{State}\in \{\mathsf{New, Running, Suspend, Terminated}\}$. 
State transitions are triggered by transactions. When a group of nodes $\mathsf{G}$ intends to establish a new OTCE, they first create a transaction with $\Vec{\mathsf{E}}=(\mathsf{EID}, \mathsf{New}, \mathsf{G}, \Delta T, \mathsf{\bot})$. The creation process is fully executed by a smart contract, which outputs $\Vec{\mathsf{E}}=(\mathsf{EID}, \mathsf{Running}, \mathsf{G}, \Delta T, \mathsf{\bot})$. The information of the registered OTCE is thereby confirmed by the smart contract. $\Delta T$ is a time duration that regulates how long $\Vec{\mathsf{E}}$ can last. Since there is no precise synchronized clocks in $\mathsf{G}$, $\Delta T$ is measured by the block height. The $\mathsf{Suspend}$ state is wakened when $\mathsf{G}$ needs to temporarily stop an application and continues later. The corresponding metadata can be stored on-chain for context switching. Once $\mathsf{G}$ decides to continue, it changes its status from $\mathsf{Suspend}$ back to $\mathsf{Running}$. The cases of suspending an OTCE should be common in a metaverse since applications may frequently need to be stopped and resumed to save resources. When computation processes are completely finished or $\Vec{\mathsf{E}}$ expires, $\Vec{\mathsf{E}}$ should be closed. The termination of $\Vec{\mathsf{E}}$ can be initiated by $\mathsf{G}$ if the task is finished before $\Vec{\mathsf{E}}$ expires, or the smart contract automatically triggers a transaction to terminate $\Vec{\mathsf{E}}$ when $\Vec{\mathsf{E}}$ expires.

\subsubsection{Security Plan According to Trust Value}
To achieve efficient resource utilization, we map trust values to different security plans in OTCE. For example, PBFT~\cite{castro1999practical} is good for a group of $N$ nodes among which at most $f=N/3$ of them can be Byzantine. A generic Paxos~\cite{lamport2001paxos} can solve consensus when at most $f=N/2$ nodes can be faulty. Since the generic Paxos has higher performance but weaker resiliency compared to PBFT, it might suffice for a group with a high trust level. Nevertheless, if the group suffers from Byzantine attacks, it needs PBFT instead. Mapping trust levels to security plans benefits resource allocation and thus can enhance efficiency. The mapping can be denoted by $\Vec{TV}\rightarrow \Vec{SP}$, where $\Vec{TV}$ (Trust Value) and $\Vec{SP}$ (Security Plan) can both reside in high dimensional spaces. One can use a matrix to represent a linear mapping or adopt machine learning and deep learning approaches to determine non-linear mappings. 
Note that an OTCE can be dynamically adjusted according to the changing trust values. For example, if an OTCE needs to shift from the generic Paxos to PBFT due to the appearance of Byzantine nodes, it can be changed through a confirmed transaction.

\subsubsection{Blockchain Virtual Machine (BVM): Decentralized Computation over Decentralized Data}\label{sec:BVM}
Trusted computing is a necessity in a metaverse since the society of the metaverse cannot be long live without establishing trust. If computational results are unreliable and unpredictable, no one knows what would happen next and thus the metaverse is in chaos. IEEE has defined ``trust'' as the level of predictability. In our architecture, we need all computational results to be fault-tolerant and verifiable. Specifically, we allow some nodes to deviate from the protocols or output wrong intermediary results but require that the final results must be correct. To avoid malicious behaviors, we need the results to be verifiable so that everyone can confirm the result without incurring much computational overhead. Verifiable computation is used to outsource computational tasks from clients to workers with dishonest behaviors. In a metaverse, we require critical computation to be verifiable. The approaches to realize verifiable computation include trusted hardware (e.g., Intel SGX and ARM Trustzone), secure multi-party computation (MPC), commitment, and zero-knowledge proofs~\cite{yu2017survey}, which can be applied to metaverses based on specific needs. Fault-tolerant and verifiable computing replaces centralized trusted third parties in a metaverse. 

Compared to the Ethereum Virtual Machine (EVM), the blockchain virtual machine (BVM) used in our architecture should support a decentralized execution of smart contracts. In an EVM, each miner executes the entire smart contract on its own and competes with other independent miners. In such a competition, only one miner can win, and the others finally lose, which wastes a huge amount of resources. By using BVM, nodes can collaboratively execute a smart contract, which decreases the execution time and also avoids undesirable race conditions. To ensure safety and liveness, task allocation follows a directed acyclic graph (DAG) representation of smart contracts.

A decentralized storage network (DSN) contributes to building a decentralized storage market without relying on cloud storage. In our architecture, decentralized computation and decentralized storage are merged into a single system to provide basic computing/storage functions for a metaverse. Recall that BVMs collaborate on contract execution. With a DSN, each BVM can compute on nearby or local data without querying big data from remote servers. By this way, the computation can be more real-time and data owners can have stronger privacy guarantees compared to traditional approaches.












\section{Conclusion}
In this article, we present a trustless blockchain-enabled metaverse architecture. The architecture provides an efficient coordination method of software and hardware. To improve performance while preserving security, we propose a few new techniques, including DID attestation and trust evaluation by hypergraph, OTCE, and BVM. In our future study, we intend to instantiate our proposed metaverse architecture by building a demo metaverse.

\printcredits

\bibliographystyle{cas-model2-names}
\bibliography{references}

\end{document}